\documentstyle[aps,12pt,preprint]{revtex}
\begin{document}
\preprint{RU--95--19}
\title{Exponentiation of multiparticle amplitudes in scalar theories.\\
II. Universality of the exponent}

\author{M. V. Libanov$^1$, D. T. Son$^{1,2}$, and S. V. Troitsky$^1$}
\address{$^1$ Institute for Nuclear Research of the Russian Academy of
Sciences,\\
         60th October Anniversary Prospect 7a, Moscow 117312 Russia\\
         $^2$ Department of Physics and Astronomy, Rutgers University,\\
         Piscataway, New Jersey 08855--0849 USA}

\date{March 1994}
\maketitle
\begin{abstract}
It has been shown recently that the amplitude of the creation of $n$
real scalar particles by one virtual boson near $n$--particle
threshold exhibits exponential behavior at $n\sim1/\lambda$. We extend
this result to the processes of multiparticle production at threshold
by two virtual bosons. We find that both the tree--level amplitude and
leading--$n$ loop corrections have the same exponential behavior, with
the common exponent for $1\to n$ and $2\to n$ processes, for various
kinematics. This result strongly indicates that the exponent for
multiparticle amplitudes is independent of the initial state and that
there may exist a semiclassical approach to the study of multiparticle
production.
\end{abstract}

\newpage

\section{Introduction}

A reliable calculation of multiparticle amplitudes remains an
interesting problem in quantum field theory. The problem exists in
most weakly coupled theories
\cite{Cornwall,Goldberg,Voloshin,Brown,AKP,Smith,LRT1,Mak1,Mak2}, including
the simplest case of ${\lambda\over4}\varphi^4$ model, where it has been
observed that the tree
amplitude to produce $n$ final particles from an initial virtual one
exhibits a factorial dependence on the number of outgoing particles like
$n!\lambda^{n/2}$, which leads to an unacceptably rapid growth of the
transition rate at the tree level at large enough $n$,
$n\sim1/\lambda$. Since the contribution from the first loop is of
order $\lambda n^2$ as compared to the tree--level result, and from
the $k$-th loop one expects a contribution of order $(\lambda n^2)^k$,
it is clear that the calculation of the amplitude at $n\sim1/\lambda$
requires the summation of the whole perturbation series. Recently, it
has been demonstrated that at the $n$-particle threshold, leading in
$n$ contributions from each loop sum up to exponent \cite{LRST}
\begin{equation}
  A_{1\to n}=A^{\text{tree}}_{1\to n}\cdot\text{e}^{B\lambda n^2}
  \label{i}
\end{equation}
where $B$ is some numerical constant and
\begin{equation}
  A^{\text{tree}}_{1\to n}=n!\left({\lambda\over 8}\right)^{n-1\over2}
  \label{1ntree}
\end{equation}
is the tree amplitude (the scalar boson mass is set equal to 1).
Equation (\ref{i}) determines correctly the threshold amplitude with
the account of all loops at $n\sim1/\sqrt{\lambda}$, in which case the
loops have the same order of magnitude as the tree--level
contribution. More importantly, at $n\sim1/\lambda$, Eq.\ (\ref{i})
provides strong indication for the exponentiation of the loop
corrections. In other words, one expects that at $n\sim1/\lambda$, the
threshold amplitude has the following form,
\begin{equation} A(1\to n)\cong\sqrt{n!}
  \cdot\text{e}^{{1\over\lambda}F(\lambda n)}
  \label{i*}
\end{equation}
where $F$ is some function that has the following expansion at small
$\lambda n$,
\[
  F(\lambda n)=F_{\text{tree}}+B\lambda^2 n^2 + O(\lambda^3 n^3)
\]
with
\begin{equation}
  F_{\text{tree}}={\lambda n\over2}\ln{\lambda n\over8}-{\lambda n\over2}
  \label{*}
\end{equation}
so at $\lambda n\ll1$, Eq.\ (\ref{i*}) reduces to Eq.\ (\ref{i}). The
behavior of the function $F$ is presently unknown at $\lambda n\sim1$.

It is likely that the exponential form of the amplitude, Eq.\
(\ref{i*}), is a consequence of the semiclassical nature of the
processes of multiparticle production. So, one may hope that there
exists some semiclassical--type technique for calculating the
amplitudes of these processes.

In this direction, one interesting suggestion
\cite{Voloshin,Khlebnikov,GorVol} is to try to generalize the
Landau method for the calculation of semiclassical matrix elements from
one--dimensional quantum mechanics to field theory. The Landau method
\cite{Landau} is a powerful technique which allows one to calculate the
matrix elements of almost any regular operator between two
semiclassical states, with different energies, of a particle moving in
one--dimensional potential. In field theory, the direct
analogues of these matrix elements are the multiparticle amplitudes:
for instance, the amplitude to produce $n$ scalar bosons at threshold by
a virtual particle can be written in the form $\langle n|\phi|0\rangle$,
where $|n\rangle$ is the state with $n$ particles at threshold.

One specific feature of the semiclassical matrix elements in
one--dimensional quantum mechanics which is captured by the Landau
method is that they do not depend, to the exponential precision, on
the operator for which they are calculated; they depend only on the
states between which the operator is sandwiched, as well as on the
details of the potential. Therefore, if there really exists a
generalization of the Landau method to field theory, one would expect
that the exponents of the matrix elements $\langle n|\hat{O}|0\rangle$
should not depend on the operator $\hat{O}$, provided the latter does
not depend on the coupling constant $\lambda$. In other words, the
multiparticle amplitudes should be the same, to the exponential
precision, for all few--particle initial states. So, a check of the
existence of the Landau--type procedure for calculating multiparticle
amplitudes will be the demonstration that they are indeed independent
of the initial state. Note that similar conjecture has been made for
instanton--induced processes \cite{RubTin,Tin}.

Presently, it is unclear how to perform reliable calculations in the
regime $\lambda n\sim1$. However, the technique of Ref.\ \cite{LRST}
can be generalized to deal with initial states that may contain more
than one virtual particle. In this way one should be able to verify
that the exponential part of the tree level amplitudes is the same as
in Eq.\ (\ref{*}) for all few--particle (containing much less than
$1/\lambda$ particles) initial states, and that the leading--$n$ loop
corrections exponentiate to $\exp(B\lambda n^2)$ with exactly the same
coefficient $B$ as in Eq.\ (\ref{i}).

In this paper we perform this check for initial states that contain
two virtual bosons and final states containing $n$ real bosons at rest
(${\bf p}_f=0$). Since there are two incoming particles, the energy
can be distributed arbitrarily between them. The first case that we
consider is the amplitude $2\to n$ integrated over 4--momentum of one
initial particle. This quantity is equal to the matrix element
$\langle n|\varphi^2|0\rangle$ at $n$--particle threshold. In two
other cases we evaluate the amplitude of scattering of two particles
with 4--momenta $(E,{\bf p})$ and $(n-E,{\bf -p})$ into $n$ particles
at rest (we set the mass of the boson equal to 1) in two different
regimes: at $E\ll n$, so that the energy of one initial particle is
much smaller than that of the other, and at $E$ of order $n$ when the
energy of the two initial particles are comparable to each other (we
will take the spatial momentum ${\bf p}$ of the incoming particle to
be small, $|{\bf p}|\sim 1$). In the latter case we consider the
regime when $E$ does not belong to the interval $(0,n)$, so one
initial particle carries negative energy (it may appear more natural
to view this particle as an outgoing one; this is of course a matter
of terminology: this particle is virtual anyway, and we are free to call
it incoming). The interval $0<E<n$ is peculiar: for these values of
$E$ even the tree--level amplitude has singular behavior as a function
of $E$, which is a consequence of the threshold kinematics, and does
not have a well defined limit in the large $n$, fixed $E/n$ regime.
Due to this peculiarity, the case $0<E<n$ is not suitable for our main
purpose.

The technique that we will apply is a direct extension of one
developed in Ref.\ \cite{LRST}. In all three cases, the calculations
show that the tree level amplitude and the leading-$n$ loop
corrections exponentiate to the same exponent as in the case of $1\to
n$ process. This result strongly indicates that the exponent of
multiparticle amplitudes in fact do not depend on the initial state
and supports the existence of some semiclassical, possibly
Landau--type, approach to the study of multiparticle production.

The paper is organized as follows. In Section \ref{GF} we will briefly
review the approach developed in Ref.\ \cite{LRST} for summing
leading--$n$ loop corrections. In Section \ref{phi2}, this technique
is applied for calculating matrix elements of the operators
$\varphi^2$.  Section \ref{epsilon} is devoted to $2\to n$ amplitudes
in the case of one soft initial particle, while in Section
\ref{epsilonn} the case of two hard initial particles is considered.
Section \ref{Conclusion} contains concluding remarks. Appendix is
devoted to a brief discussion of the case $0<E<n$, and the tree and
one--loop results are presented.

\section{General Formalism}
\label{GF}

We consider $\varphi^4$ theory with unbroken discrete symmetry in
$(d+1)$ dimensional spacetime with the Lagrangian
\begin{equation}
L=\frac{1}{2}(\partial_\mu\varphi)^2-\frac{m^2}{2}\varphi^2-\frac{\lambda}{4}
\varphi^{4}
\label{0}
\end{equation}
Hereafter we set the mass of the boson $m$ to be equal to 1.

For calculating multiparticle amplitudes at threshold, there exists a
convenient formalism \cite{Brown} which reduces the problem to the
calculation of Feynman graphs in certain classical background. In
Ref.\ \cite{Brown} this formalism has been developed for the $1\to n$
amplitude, but it is easy to extend it for treating processes with two
incoming particles.

Let us outline briefly this technique (see Ref.\ \cite{Brown} for
details).  Consider a transition from an initial virtual particle with
$(d+1)$--momentum $P_{\mu}=(n,{\bf 0})$ into $n$ final particles, each
with $(d+1)$--momentum $(1,{\bf 0})$. The reduction formula for the
amplitude can be written in the following form,
\begin{equation}
  \langle n|\phi(x)|0\rangle=\lim_{\varepsilon_0^2\to 1}\lim_{J_0\to 0}
  (\varepsilon_0^2-1)^n
      {\partial^n\over\partial J_0^n}
      \langle 0|\varphi(x)|0\rangle_{J=J_0\exp(i\varepsilon_0 t)}
  \label{recursion}
\end{equation}
where the matrix element is calculated in the presence of a source
$J=J_0 \exp{(i\varepsilon_0 t)}$.

Taking the limits $\varepsilon_0\to 0$, $J_0\to 0$ simultaneously, one
can show that $A_n$ is determined by
\begin{equation}
  A_n = {\partial^n\over\partial z_0^n} \langle 0|\varphi|0\rangle
  \left.\right|_{z=0}
  \label{brown}
\end{equation}
where the expectation value $\langle0|\varphi|0\rangle$ is calculated
in the following classical background
\begin{equation}
  \varphi_0(t) = {z_0\text{e}^{it}\over 1-{\lambda\over8}z_0\text{e}^{2it}}
  \label{background}
\end{equation}
which is a solution to the field equation.

It is straightforward to generalize this technique to the case of
matrix elements of arbitrary operators. In particular, the matrix
element of the operator $\varphi^2$ can be obtained by differentiating
its vacuum expectation value in the presence of the background field
$varphi_0$,
\begin{equation}
  \langle n |\varphi^2 | 0 \rangle =
  {\partial^n\over\partial z_0^n} \langle 0|\varphi^2|0\rangle
  \left.\right|_{z_0=0}
\label{16}
\end{equation}
Analogously, to calculate the amplitude of scattering of two initial
particles with momenta $(E,{\bf p})$ and $(n-E,-{\bf p})$ into $n$
bosons at threshold, one should differentiate the corresponding full
propagator $n$ times with respect to $z_0$,
\begin{equation}
  A_{2\to n}(E,{\bf p})=
  {\partial^n\over\partial z_0^n}
  \int\! d^{d+1}x\,d^{d+1}y\, \text{e}^{iEx^0+i(n-E)y^0}
  \text{e}^{i {\bf p}({\bf y}-{\bf x})}{\cal D}(x,y)
 \label{2}
\end{equation}
where ${\cal D}(x,y)$ is the two--point Green function calculated in
the classical background $\varphi_0$. It is convenient to use the
mixed coordinate--momentum representation,
\[
  {\cal D}_{\bf p}(x^0,y^0)\equiv \int\! d^d{\bf x}\,d^d{\bf y}\,\text{e}^{i
  {\bf p}({\bf y}-{\bf x})}{\cal D}(x,y)
\]
so Eq.\ (\ref{2}) can be rewritten in the following form,
\begin{equation}
  A_{2\to n}(E,{\bf p})=
  {\partial^n\over\partial
  z_0^n}\int\!dx^0\,dy^0\,\text{e}^{iEx^0+i(n-E)y^0} {\cal D}_{\bf
  p}(x^0.y^0).  \label{2new} \end{equation}

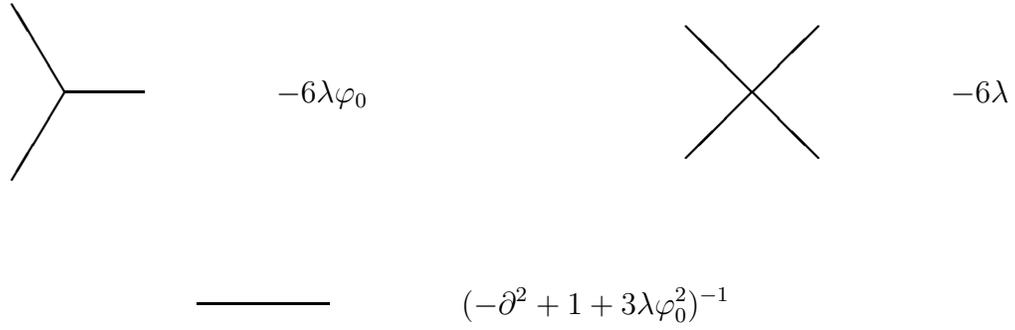
\begin{figure}
\thicklines
\begin{picture}(432,100)(0,310)
\put(100,250){\line(1,0){50}}
\put(200,246){$\displaystyle{(-\partial^2+1+3\lambda\varphi_0^2)^{-1}}$}
\put(50,330){\line(1,0){30}}
\put(50,330){\line(-3,-5){20}}
\put(50,330){\line(-3,5){20}}
\put(130,326){$\displaystyle{-6\lambda\varphi_0}$}
\put(310,330){\line(1,1){25}}
\put(310,330){\line(1,-1){25}}
\put(310,330){\line(-1,1){25}}
\put(310,330){\line(-1,-1){25}}
\put(385,326){$\displaystyle{-6\lambda}$}
\put(0,0){~}
\end{picture}

\vspace{3.5cm}

\caption{Feynman rules for calculating multiparticle amplitudes}
\label{fig:feynman}
\end{figure}

So, the amplitudes to produce $n$ final particles at threshold can be
obtained by differentiating the corresponding Green functions
(``generating functions''), calculated in the presence of the
background field $\varphi_0$, $n$ times with respect to the parameter of
the background, $z_0$. The Green functions which enter the right hand
sides of Eqs.\ (\ref{16}) and (\ref {2new}), in their turn, can be
computed in the perturbation theory around the classical background
$\varphi_0$. It is convenient to introduce the Euclidean time
variable,
\begin{equation}
  \tau=it+\frac{1}{2}\ln\frac{\lambda}{8}+\ln z_0
 +\frac{i\pi}{2},
  \label{tau}
\end{equation}
in terms of which the background field has the form
\[
  \varphi_0(\tau)=-i\sqrt{\frac{2}{\lambda}}\frac{1}{\cosh \tau}.
\]
Note that the background field has a singularity at $\tau=i\pi/2$. By
expanding the Lagrangian around the background $\varphi_0$ one obtains
the Feynman rules shown in Fig. \ref{fig:feynman}.

Obviously, perturbative calculations become more and more complicated
at higher loops. However, if one is interested only in the leading--$n$
contribution from each loop level, considerable simplification occurs
\cite{LRST}, which allows for the summation of the whole perturbative
series. The key point is that, at each loop level, the large--$n$
behavior of the amplitude depends only on the structure of the
singularity at $\tau=i\pi/2$ of the generating functions (in our case
$\langle0|\varphi^2|0\rangle$ or ${\cal D}_{\bf p}(x^0,y^0)$). Thus,
our strategy is to find the generating function near the singularity
and after that recover the multiparticle amplitude.

In the case of $1\to n$ process, the procedure has been developed
\cite{LRST} for obtaining the leading singularity of the corresponding
generating function at $\tau=i\pi/2$ at any given loop order, which
reduces the problem to the calculation of tree graphs in some
effective theory. We will apply this technique to the case of the
processes with two initial particles. Let us begin with the matrix
element of $\varphi^2$.

\section{Matrix elements of $\varphi^2$}
\label{phi2}

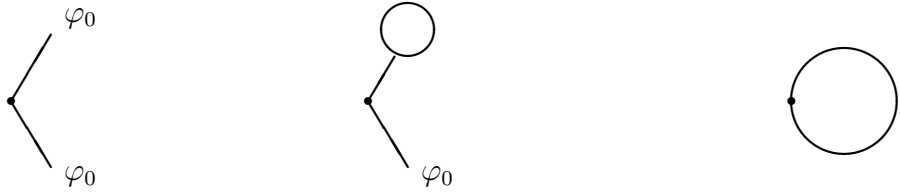
\begin{figure}
\begin{picture}(440,100)(-35,300)
\thicklines
\put(5,350){\circle*{2.5}}
\put(5,350){\line(3,5){15}}
\put(5,350){\line(3,-5){15}}
\put(25,380){$\displaystyle{\varphi_0}$}
\put(25,320){$\displaystyle{\varphi_0}$}
%\put(65,346){$\displaystyle{;}$}

\put(140,350){\circle*{2.5}}
\put(140,350){\line(3,5){10}}
\put(140,350){\line(3,-5){15}}
\put(155,377){\circle{20}}
\put(160,320){$\displaystyle{\varphi_0}$}
%\put(180,346){$\displaystyle{=}$}

%\put(210,350){\circle*{2.5}}
%\put(210,350){\line(3,5){20}}
%\put(223,372){\line(3,-5){7}}
%\put(230,383){\circle*{5}}
%\put(230,361){\circle*{5}}
%\put(210,350){\line(3,-5){15}}
%\put(230,320){$\displaystyle{\varphi_0}$}
%\put(270,346){$\displaystyle{;}$}

\put(300,350){\circle*{2.5}}
\put(320,350){\circle{40}}
%\put(350,346){$\displaystyle{=}$}

%\put(370,350){\circle*{2.5}}
%\put(370,350){\line(3,5){20}}
%\put(370,350){\line(3,-5){20}}
%\put(390,383){\circle*{5}}
%\put(390,317){\circle*{5}}
\end{picture}
\caption{The tree and one--loop graphs which contribute to the matrix
element $\langle0|\varphi^2|0\rangle$.}
\label{fig:max}
\end{figure}

\begin{figure}
\begin{picture}(432,60)(0,250)
\thicklines
\put(105,326){$\displaystyle{\frac{1}{4}}$}
\put(140,330){\circle{40}}
\put(140,310){\line(0,1){40}}
\put(120,330){\circle*{2.5}}
\put(175,326){$\displaystyle{=}$}
\put(121,345){\scriptsize{1}}
\put(121,310){\scriptsize{2}}
\put(141,328){\scriptsize{3}}
\put(154,345){\scriptsize{4}}
\put(140,280){a}

\put(200,326){$\displaystyle{\frac{1}{4}}$}
\put(215,330){\circle*{2.5}}
\put(215,330){\line(3,-5){20}}
\put(215,330){\line(3,5){20}}
\put(231,354){\line(3,-5){6}}
\put(231,306){\line(3,5){6}}
\put(235,364){\circle*{5}}
\put(235,344){\circle*{5}}
\put(235,315){\circle*{5}}
\put(235,295){\circle*{5}}
\put(275,280){b}

\put(250,326){$\displaystyle{+}$}
\put(275,330){\circle*{2.5}}
\put(275,330){\line(1,0){40}}
\put(315,330){\line(3,5){10}}
\put(315,330){\line(3,-5){10}}
\put(326,350){\circle*{5}}
\put(326,310){\circle*{5}}
\put(275,330){\line(3,5){10}}
\put(295,330){\line(3,5){10}}
\put(286,350){\circle*{5}}
\put(306,350){\circle*{5}}
\end{picture}

\caption{The two--loop graph for the operator $\varphi^2$ (a) and a
representation (b) of its leading singularity}
\label{fig:phi2}
\end{figure}
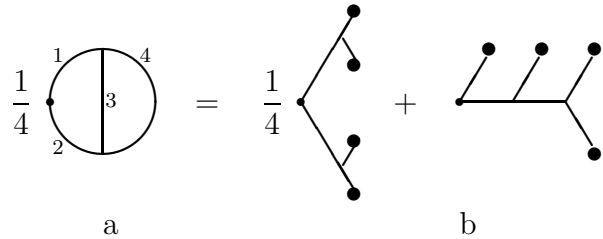

Let us first discuss the Feynman graphs that contribute to the
generating function, $\langle0|\varphi^2|0\rangle$. There is one
tree--level graph shown in Fig.\ \ref{fig:max}. At the one--loop
level, there are two different types of graphs. The two--loop graphs
are much more numerous, only one of them is presented in Fig.\
\ref{fig:phi2}a.

As explained above, at each loop level it is sufficient to calculate
only the leading singularity of the generating function. The latter
can be found by extending the technique of Ref.\ \cite{LRST}, that
reduces the problem to the evaluation of tree graphs. Since this
extension is straightforward, let us summarize here only the
prescription (see Ref.\ \cite{LRST} for details):
\begin{itemize}
\item For a given loop graph one should cut some propagator lines in
such a way that the resulting graph is tree and connected. In
cases when there are various ways to cut, the result is given by the
sum over all possibilities. The number of lines to be cut is equal to
the number of loops.
\item For each propagator line, say, $G(\tau,\tau')$, that has been
cut, one attributes the factors $B^{1/2}\lambda\varphi^2(\tau)$ and
$B^{1/2}\lambda\varphi^2(\tau')$ to the two vertices that this line
connects, i.e., at $\tau$ and $\tau'$, respectively. The constant $B$
is defined by
\[
  B=\int\!{d^d{\bf p}\over(2\pi)^d}\,{9\over8\omega_{\bf p}
  (\omega_{\bf p}^2-1)(\omega_{\bf p}^2-4)}
\]
where $\omega_{\bf p}=\sqrt{{\bf p}^2+1}$. All propagators that have
not been cut should be replaced by the operator
$(\partial_{\tau}^2+3\lambda\varphi_0^2)^{-1}$. Therefore, the tree
graphs obtained by our cutting procedure do not contain spatial
momenta explicitly.
\end{itemize}

Let us demonstrate how this prescription works for a particular graph
shown in Fig.\ \ref{fig:phi2}a. This graph has a symmetry factor of
$1/4$. Since it is a two--loop graph, the number of propagators to be
cut is 2. There are in fact 5 possibilities to cut in such a way that
the graph remains connected: one can cut the lines (1,3), (2,3),
(1,4), (2,4), or (3,4) (one cannot cut two propagators (1,2), since
then the graph will become disconnected).  In fact, one can see that
the first four ways lead to essentially the same graphs, while the
graph obtained in the fifth case has another topology. As a result,
one obtains a sum of two graphs with symmetry factors of 1 and $1/4$,
as shown in Fig.\ \ref{fig:phi2}b.  The black circles (``bullets'')
represent the factors $B^{1/2}\lambda\varphi^2$.

By analyzing the tree graphs obtained by cutting the initial loop
graphs, Figs.\ \ref{fig:max} and \ref{fig:phi2}, one can see that
they have the form of a tree cascade starting from two initial lines
corresponding to the two operators $\varphi(\tau)$ (a two--branch
tree). Each branch contains contributions of $\varphi_0$,
$B^{1/2}\lambda\varphi_0^2$ and higher power of $\lambda$. One observes
(analogously to the $1\to n$ case) that the same series of graphs is
obtained when one solves the field equation without the mass term,
\begin{equation}
  \partial_{\tau}^2\varphi+\lambda\varphi^3=0
  \label{classical}
\end{equation}
perturbatively with respect to $\lambda$, with the condition that the
 first two terms in the expansion over $\lambda$ are
$\varphi_0+B^{1/2}\lambda\varphi_0^2$.

Eq.\ (\ref{classical}) has the following  exact solution
\begin{equation}
  \varphi_{\text{cl}}=-\sqrt{2\over\lambda}\,{1\over\tau-\tau_0}
  \label{tau0}
\end{equation}
which has a single pole at $\tau=\tau_0$. One can fix $\tau_0$ by
requiring that the solution has the expansion
$\varphi_{\text{cl}}=\varphi_0+B^{1/2}\lambda\varphi_0^2+\cdots$. As a
result, one obtains
\[
  \tau_0=i\pi/2-\sqrt{2B\lambda}.
\]
Then
$\varphi_{\text{cl}}$ can be represented in the following form,
\begin{equation}
  \varphi_{\text{cl}}=\varphi_0\sum_k (B^{1/2}\lambda\varphi_0)^k
  \label{phicl}
\end{equation}

At first sight, the generating function is equal to
$\varphi_{\text{cl}}^2$, which has the following expansion,
\begin{equation}
  \varphi_{\text{cl}}^2   =\varphi_0^2\sum_{k=0}^\infty(k+1)
  (B^{1/2}\lambda\varphi_0)^k
\label{11*}
\end{equation}
However, this is not true. First, one notes that all graphs obtained
by the cutting procedure contain even number of bullets, so there
should be only even powers of $B^{1/2}\lambda$ in the series for the
generating function. Second, from every tree graph with $2l$ bullets
one can reconstruct $(2l)!/(2^ll!)$ graphs of the original theory at
$l$--loop order by pairing the bullets into propagator lines. So, to
recover the generating function one should omit in Eq.\ (\ref{11*})
all terms with odd $k$ and for terms with even $k$, $k=2l$, one should
multiply the coefficient by the factor $(2l)!/(2^ll!)$. In this way
one obtains,
\begin{equation}
  \langle0|\varphi^2|0\rangle=\varphi^2_0\sum_{l=0}^\infty
  \frac{(2l+1)!}{2^ll!}(\lambda^2B\varphi_0^2)^k
\label{12}
\end{equation}
To obtain the matrix element $\langle n|\varphi^2|0\rangle$ one
substitutes Eq.\ (\ref{background}) into Eq.\ (\ref{12}) and
differentiates $n$ times with respect to $z$. Recalling Eq.\
(\ref{tau}), one obtains the following result,
\[
  \langle 0|\varphi^2|n\rangle=\frac{n!n}{2}\left(\frac{\lambda}{8}\right)^{
  \frac{n}{2}-1}\sum_{k=0}^\infty\frac{(\lambda
  Bn^2)^k}{k!}=A_{\varphi^2}^{\text{tree}}\text{e}^{B\lambda n^2}
\]
where
  \[A_{\varphi^2}^{\text{tree}}=\frac{n!n}{2}\left(\frac{\lambda}{8}
  \right)^{\frac{n}{2}-1}
\]
can be identified with the matrix element calculated at the tree
level, which differs from the tree $1\to n$ amplitude (\ref{1ntree})
only by a pre--exponential factor proportional to $n$.  The
leading--$n$ loop contributions exponentiate to the same factor of
$\exp(B\lambda n^2)$. Therefore, the exponent for the matrix element
of $\varphi^2$ coincides with the $1\to n$ amplitude, which is the
desired result.

\section{Amplitudes $2\to \lowercase{n}$ with one soft initial particle}
\label{epsilon}

In this section we consider the process of scattering of two initial
virtual particles, among which one is soft (with energy $E\ll n$) and
the other is hard, into $n$ final particles at threshold (the spatial
momenta of initial particles are assumed to be small, i.e. of order
1).

It is convenient to rewrite Eq.\ (\ref{2new}) in a slightly modified
form. One notes that it is sufficient to consider only one integral
over $dy^0$, the integral over $dx^0$ results in the delta--function
of energy conservation, which we will not explicitly write in what
follows. Furthermore, one introduces the notation
$z(t)=z_0\text{e}^{it}$, and goes to Euclidean time, then the $2\to n$
amplitude is written as follows,
\[
  A_{2\to n}(E,{\bf p})={\partial^n\over\partial z^n} G(\tau)
\]
where
\begin{equation}
  G(\tau)=\text{e}^{E\tau}\int\!d\tau'\,{\cal D}_{\bf p}(\tau,\tau')
  \text{e}^{-E\tau'}
  \label{G2n}
\end{equation}
We will call $G(\tau)$ the generating function for the $2\to n$
amplitude.  At the tree level this result was obtained firstly in
\cite{Zeros} by direct summation of graphs. Let us begin with discussing
the tree level.

\subsection{Tree level}

At the tree level, it is easier to find the amplitude and then recover
the generating function. The tree--level amplitude can be found
directly from the exact tree propagator in the background field
$\varphi_0$, Eq.\ (\ref{background}) \cite{prop1,prop2}
\[
  D_{\bf p}(\tau,\tau')=
  {1\over W_{\bf p}}(f_1^{\omega}(\tau)f_2^{\omega}(\tau')
  \theta(\tau'-\tau)+f_2^{\omega}(\tau)f_1^{\omega}(\tau')
  \theta(\tau-\tau'))
\]
where
\[
  W_{\bf p}=2\omega(\omega^2-1)(\omega^2-4),\qquad
  \omega=\sqrt{{\bf p}^2+1}
\]
and
\[
  f_1^{\omega}(\tau)=\text{e}^{\omega\tau}
  \left({12\over(1+\text{e}^{2\tau})^2}+
  {6(\omega-2)\over1+\text{e}^{2\tau}}+(\omega-1)(\omega-2)\right)
\]
\[
  f_2^{\omega}(\tau)=f_1^{-\omega}(\tau)
\]
To evaluate the integral (\ref{G2n}), one expands the functions $f_1$
and $f_2$ in series,
\[
  f_1^\omega(\tau)=\text{e}^{\omega\tau}\sum_{k=0}^\infty(-1)^kf_{1k}\text
  {e}^{2k\tau}
\]
\begin{equation}
  f_2^\omega(\tau)=\text{e}^{-\omega\tau}\sum_{k=0}^\infty(-1)^kf_{2k}\text
  {e}^{2k\tau}
\label{14}
\end{equation}
where
\[
  f_{1k}=\delta_{k0}(\omega-1)(\omega-2)+6\omega+12k
\]
\[
  f_{2k}=\delta_{k0}(\omega+1)(\omega+2)-6\omega+12k
\]
The tree amplitude $2\to n$ can be now obtained by direct calculation.
One finds,
\begin{equation}
  A^{\text{tree}}_{2\to n}(E, {\bf p})=
  n!\left({\lambda\over8}\right)^{n/2}{1\over W_{\bf p}}\sum_{k=0}^{k=n/2}
  \left({f_{1k}f_{2({n\over2}-k)}\over2k+\omega-E}-
  {f_{2k}f_{1({n\over2}-k)}\over2k-\omega-E}\right)
  \label{2nsum}
\end{equation}
It is worth noting that $A^{\text{tree}}_{2\to n}(E,{\bf p})$ develops
a series of poles at $E=2k\pm\omega$. The physical reason for these
poles is that for these values of $E$ the tree graphs may include a
propagator with on--shell momentum (see Fig.\ \ref{fig:tree2n}).

To obtain the asymptotic behavior of the amplitude at large $n$, one
notes that at $E\sim1$, the sum in Eq.\ (\ref{2nsum}) is saturated by
a finite number of terms with $k\sim1$. So we can replace
$f_{1,2(n/2-k)}$ by $6n$ and extend the sum to $k=\infty$. We obtain,
\[
  A^{\text{tree}}_{2\to n}(E, {\bf p})=
  6n!n\left({\lambda\over8}\right)^{n/2}{1\over W_{\bf p}}
  \sum_{k=0}^{\infty}\left(
  {f_{1k}\over2k+\omega-E}-{f_{2k}\over2k-\omega-E}\right)
\]
Note that the sum in this equation converges and does not depend on
$n$. Let us introduce the function,
\[
  C(E,{\bf p})={3\over2}{1\over W_{\bf p}}\sum_{k=0}^{\infty}\left(
  {f_{1k}\over2k+\omega-E}-{f_{2k}\over2k-\omega-E}\right)
\]
The tree amplitude can now be rewritten in the form
\begin{equation}
  A^{\text{tree}}_{2\to n}(E, {\bf p})=
  4n\cdot n!\left({\lambda\over8}\right)^{n/2}C(E,{\bf p})
  \label{Atree}
\end{equation}

Note that the tree amplitude is proportional to the tree matrix
element of $\varphi^2$,
\[
  A^{\text{tree}}_{2\to n}(E, {\bf p})=
\lambda C(E,{\bf p})\langle n|\varphi^2|0\rangle_{\text{tree}}
\]
and correspondingly differs from the tree $1\to n$ amplitude
(\ref{1ntree}) only by a factor of $n$. Therefore the generating
function is proportional to that of the operator $\varphi^2$,
\begin{equation}
  G^{\text{tree}}(\tau)=\lambda C(E,{\bf p})\varphi_0^2(\tau)
  \label{15*}
\end{equation}
One can verify by differentiation that the generating function
(\ref{15*}) indeed gives rise to the amplitudes (\ref{Atree}) at large
$n$.

\subsection{Loop corrections}

To sum leading--$n$ loop corrections to the $2\to n$ amplitude, one
should calculate the generating function
\[
  G(\tau)=\text{e}^{E\tau}\int\!d\tau'{\cal D}(\tau,\tau')\text{e}^{-E\tau'}
\]
in the region close to singularity. The technique that we apply here
is the same as that used in the case of the operator $\varphi^2$: one
cuts every loop graph so that it becomes tree and connected and
attaches bullets to the cut lines. In this way one obtains the leading
singularity of the propagator $\cal{D}(\tau,\tau')$ as a sum of graphs
that typically have the form shown in Fig.\ \ref{fig:smallE}. One can
see that these graphs are the same as those obtained by expanding the
propagator in the background field $\varphi_{\text{cl}}$
\[
  D_{\text{cl}}(\tau,\tau')=
  (-\partial_{\mu}^2+1+3\lambda\phi_{\text{cl}})^{-1}
\]
in the perturbation series in $\lambda$, where the background field
$\varphi_{\text{cl}}$ and its expansion are given by Eqs.\ (\ref{tau0})
and (\ref{phicl}), respectively. The only difference is that only
graphs with even number of bullets is present for ${\cal D}$ and the
symmetry coefficients differ by a factor of $(2l)!/(2^ll!)$, where
$2l$ is the number of bullets.

The quantity
\begin{equation}
  G_{\text{cl}}(\tau)=\text{e}^{E\tau}\int\!d\tau'
  D_{\text{cl}}(\tau,\tau')\text{e}^{-E\tau'}
  \label{Gcl}
\end{equation}
can be easily calculated since the fields $\varphi_{\text{cl}}$ and
$\varphi_0$ have the same behavior around their singularities, while
the singularities are located at different points.  Recalling Eq.\
(\ref{15*}) one writes
\[
  G_{\text{cl}}(\tau)=\lambda C(E,{\bf p})\varphi^2_{\text{cl}}(\tau)
\]
\begin{equation}
  =\lambda C(E,{\bf p})\varphi^2_0\sum_{l=0}^\infty(k+1)
  (\lambda^2B\varphi_0^2)^k
\label{12new}
\end{equation}
$G(\tau)$ can be obtained from this expression by dropping all terms
in the sum with odd $k$ and correcting the coefficients by the factors
$(2l)!/(2^ll!)$. Comparing with Eq.\ (\ref{12}), one finds that the
generating function in our case is proportional to that of the matrix
element of $\varphi^2$,
\[
  G(\tau)=\lambda C(E,{\bf p})\langle0|\varphi^2|0\rangle
\]
so the loop corrections exponentiate in the same way as for
$\langle n|\varphi^2|0\rangle$,
\[
  A_{2\to n}(E,{\bf p})=A^{\text{tree}}_{2\to n}(E,{\bf p})
  \text{e}^{B\lambda n^2}
\]

So, we have established that both the tree expression and the
leading--$n$ loop corrections for the amplitude $2\to n$ in the case
when one initial particle is soft sum up to
$\exp({1\over\lambda}F_{\text{tree}}+B\lambda n^2)$.

\section{Amplitudes $2\to \lowercase{n}$ with two hard initial particles}
\label{epsilonn}

In this section we consider the case when both initial particles have
energies of order $n$. For the amplitude to have a regular limit at
large $n$, we take the energy of one incoming particle, $E$, to be
negative, while the energy of the other is larger than $n$. The case
when energies of both initial particles are positive is briefly
discussed in Appendix.

\subsection{Tree level}

The tree amplitude in our case can be determined from its
representation in terms of the sum in Eq.\ (\ref{2nsum}). However,
from Eq.\ (\ref{2nsum}) it is not straightforward to extract its
asymptotic behavior in the regime $n\to\infty$, $E/n=\mbox{fixed}$. We
adopt here another approach. We will find the generating function from
the equation that it obeys,
\begin{equation}
  (-\partial^2_\tau+3\lambda\varphi_0^2)
  \left(\text{e}^{-E\tau}G^{\text{tree}}(\tau)\right)=
   \text{e}^{-E\tau}
\label{18}
\end{equation}
(we have omitted the term $\omega^2$ which is inessential near the
singularity). Since the amplitude to produce $n$ particles depends on
the details of the behavior of generating function in the region
$|\tau-i\pi/2|\sim1/n$, and $E\sim n$, we look for the solution of
Eq.\ (\ref{18}) in the region $|\tau-i\pi/2|\sim1/E$. Near the
singularity, Eq.\ (\ref{18}) reduces to
\begin{equation}
  \left(-\partial_{\tau}^2+{6\over(\tau-i\pi/2)^2}\right)
  \left(\text{e}^{-E\tau}G^{\text{tree}}(\tau)\right)=
  \text{e}^{-E\tau}
  \label{18new}
\end{equation}
Eq.\ (\ref{18new}) can be solved exactly, the solution has the
following form
\[
  G^{\text{tree}}(\tau)=-{1\over 5E^2}\left({6\over
  E^2(\tau-i\pi/2)^2}+{6\over
  E(\tau-i\pi/2)}+3+E(\tau-i\pi/2)-
  E^2(\tau-i\pi/2)^2-\right.
\]
\begin{equation}
  -\left. E^3(\tau-i\pi/2)^3 \text{e}^{E\tau}\int
  \limits_{-\infty}^{\tau}\!\frac{\text{e}^
  {-E\tau' } } {\tau'-i\pi/2 } \,d\tau'\right)
\label{18*}
\end{equation}
We have kept in Eq.\ (\ref{18*}) not only the leading singular term,
but all terms which are of order $E^{-2}$ at $E|\tau-\pi/2|\sim1$.
Those terms are equally important for the evaluation of the amplitude:
while the leading singular term (the first term in parenthesis in
Eq.\ (\ref{18*})) gives rise to the contribution of order $nE^{-4}$,
the second term produces $E^{-3}$, etc. (see below).  Note that $E<0$,
so the integral in this equation converges.

The large--$n$ asymptotics of the amplitude is uniquely determined by
the behavior of the generating function which has been found in Eq.\
(\ref{18*}). To recover the amplitude one should find a series in
$\text{e}^{2\tau}$ which has the same behavior around the singularity
as the r.h.s. of Eq.\ (\ref{18*}). Apparently there are various ways
to write down the series in $\text{e}^{2\tau}$ that coincides with
Eq.\ (\ref{18*}) near the singularity, but they all yield the same
result for the amplitude.  Technically, the most convenient way to
write the series is to make the following replacement,
\begin{equation}
  \tau-i\pi/2\to -\frac{1+\text{e}^{2\tau}}{2}
\label{18**}
\end{equation}
which is valid near the singularity, so the generating function
obtains the form
\[
  G^{\text{tree}}(\tau)=-{1\over 5E^2}\left({24\over E^2(1+\text{e}^{2\tau})^2}
  -{12\over E(1+\text{e}^{2\tau})}+3-{E\over2}(1+\text{e}^{2\tau})-
  {E^2\over4}(1+\text{e}^{2\tau})^2\right.
\]
\[
  \left.-{E^3\over4}(1+\text{e}^{2\tau})^3\text{e}^{E\tau}
  \int\limits_{-\infty}^{\tau}\!{\text{e}^{-E\tau'}
  \over(1+\text{e}^{2\tau})}\,d\tau'\right)
\]
which is indeed a series in $\text{e}^{2\tau}$. Recalling Eq.\ (\ref{tau}),
the tree amplitude can be derived by differentiating with respect to
$z$. One obtains,
\begin{equation}
  A_{2\to n}^{\text{tree}}=-\frac{12n!}{5}\left(\frac{\lambda}{8}\right)^
  {\frac{n}{2}}\frac{1}{E^4(n-E)^4}\left((n-E)^5
  +E^5\right)
\label{19*}
\end{equation}
Again, the exponential part of the amplitude is equal to
$\exp(\lambda^{-1}F_{\text{tree}}(\lambda n))$, while the energy
dependence enters only the pre--exponential factor.

\subsection{Loop corrections}

In analogy to the case of one soft initial particle, one should first
calculate $G_{\text{cl}}$, Eq.\ (\ref{Gcl}). To do this we recall that
$D_{\text{cl}}$ is the propagator in the background field
$\varphi_{\text{cl}}$ which has the pole at
$\tau=i\pi/2-\sqrt{2\lambda B}$. So, instead of Eq.\ (\ref{18new}) one
has
\[
  \left(-\partial^2_\tau+{6\over
  (\tau-i\pi/2+\sqrt{2\lambda B})^2}\right)\left(\text{e}^{-E\tau}
  G_{\text{cl}}
  (\tau)\right)=\text{e}^{-E\tau}
\]
Since this equation has the same form as Eq.\ (\ref{18new}), its
solution is given by a formula analogous to Eq.\ (\ref{18*}), with the
only difference that the pole at $i\pi/2$ is shifted to
$i\pi/2-\sqrt{2\lambda B}$. Making the replacement (\ref{18**}), one
obtains the solution in the following form (we do not write explicitly
the terms that are regular at the singularity)
\[
  G(\tau)=-{1\over 5E^2}\left(
  {24\over E^2\left(1+\text{e}^{2\tau}-2\sqrt{2\lambda B}\right)^2}+
  {12\over E\left(1+\text{e}^{2\tau}-2\sqrt{2\lambda B}\right)}\right.
\]
\begin{equation}
  -\left.{E^3\over4}\left(1+\text{e}^{2\tau}-2\sqrt{2\lambda B}\right)^3
  \text{e}^{E\tau'}
  \int\limits_{-\infty}^{\tau}\!\frac{\text{e}^{-E\tau' } }
  {\left(1+\text{e}^{2\tau}-2\sqrt{2\lambda B}\right)} \,d\tau'\right)
  +~\mbox{regular terms}
  \label{19**}
\end{equation}
which should be differentiated with respect to $z$ in order to obtain
the amplitude. Remembering that we should omit terms with odd powers
of $B^{1/2}$ and multiply the coefficient of $(B^{1/2})^{2l}$ by
$(2l)/(2^ll!)$, we obtain after a tedious but straightforward
calculation,
\[
  A_{2\to n}=A^{\text{tree}}_{2\to n}\sum_{k=0}^{\infty} {(\lambda Bn^2)^k
  \over k!}=A^{\text{tree}}_{2\to n}\text{e}^{\lambda Bn^2}
\]
where $A_{2\to n}^{\text{tree}}$ is given by Eq.\ (\ref{19*}).
Therefore, although in the case of two hard initial particles the
calculations are more complicated, the result is the same as in the
case of the matrix element of $\varphi^2$ or when one initial particle
is soft: leading--$n$ corrections sum up to the exponent
$\exp(B\lambda n^2)$.

\section{Conclusion}
\label{Conclusion}
We have studied the amplitude of the processes $2\to n$ in the
$\varphi^4$ theory by the technique that allows to sum up all
leading--$n$ loop corrections at $n$--particle threshold. We have
shown that the $2\to n$ amplitudes, regardless of how the initial
energy is distributed among the two initial particles (except for some
peculiar cases), coincide with the amplitude $1\to n$ to the
exponential precision, at least when only leading--$n$ loop
contributions are taken into account.  The similar result can be
easily obtained in the case of broken discrete symmetry. Our results,
though not being a rigorous proof, strongly support the hypothesis
that the amplitude of multiparticle processes is independent, in the
exponential approximation, of the initial few--particle state. The
picture here is similar to that of quantum mechanics, where the
calculation of semiclassical matrix elements by the Landau method
requires no knowledge on the details of the operator sandwiched
between the semiclassical states.  Our calculations, therefore,
indicate that there may exist an extension of the Landau method to
field theory, which, hopefully, could bring about the understanding of
multiparticle amplitudes.

\acknowledgements

The authors are indebted V.~A.~Rubakov for numerous helpful
discussions and stimulating activity. We thank Yu.~M.~Makeenko and
P.~G.~Tinyakov for valuable discussions.  D.T.S. thanks Rutgers
University for hospitality.  This work is supported in part by INTAS
grant \# INTAS-93-1630.  The work of D.T.S. is supported in part by
Russian Foundation for Fundamental Research, grant \# 93-02-3812.

\appendix
\section{$2\to \lowercase{n}$ amplitudes for $0<E<\lowercase{n}$}

In this Appendix we will briefly consider the behavior of the tree
amplitude $2\to n$ in the case when both initial particles have
positive energies of order $n$, and also present the result for
the one--loop correction to this amplitude.

The tree amplitude can be derived by making use of Eq.\ (\ref{2nsum}).
At $E\sim n$ and $0<E<n$, the sum in Eq.\ (\ref{2nsum}) is saturated
by the terms with $k\approx E/2$. For these values of $k$ one has
$f_{1,2k}\approx6E$, $f_{1,2(n/2-k)}\approx6(n-E)$. The sum in Eq.\
(\ref{2nsum}) can be extended so that $k$ runs from $-\infty$ to
$\infty$, and one writes,
\[
  A^{\text{tree}}_{2\to n}(E,{\bf p})=n!\left({\lambda\over8}\right)^{n/2}
  {36E(n-E)\over W_{\bf p}}\sum_{k=-\infty}^{\infty}\left(
  {1\over2k+\omega-E}-{1\over2k-\omega-E}\right)
\]
\[
  =n!\left({\lambda\over8}\right)^{n/2}{72E(n-E)\over\pi W_{\bf p}}
  \left(\cot{\pi\over2}(E+\omega)-\cot{\pi\over2}(E-\omega)\right)
\]
Note that the amplitude has a very rapid oscillating behavior as a
function of $E$, so it has no regular limit in the regime
$n\to\infty$, $E/n=\mbox{fixed}$, but rather the amplitude depends on,
say, the fractional part of $E/2$. Due to this behavior of the
amplitude at the tree level, one should not, in general, expect that
the loop corrections will sum up into a regular factor like
$\text{e}^{B\lambda n^2}$.

We have performed the calculation of the one--loop correction to the
amplitude by direct evaluation of the three graphs shown in Fig.\
\ref{fig:3diagrams}. At large $n$, the result is
\[
  A^{\text{1-loop}}=n!\left({\lambda\over8}\right)^{n/2}
  {18\over\pi}\lambda E(n-E)\left[
  {4B\over W_{\bf p}}\left(E^2+(n-E)^2\right)
  \left(\cot{\pi\over2}(E+\omega)-\cot{\pi\over2}(E-\omega)\right)+\right.
\]
\begin{equation}
  +\left. 9E(n-E)\int\!
  {d{\bf k}\over W_{\bf k}W_{\bf p-k}}
  \left(\cot{\pi\over2}(E+\omega_{\bf k}+\omega_{\bf p-k})-
  \cot{\pi\over2}(E-\omega_{\bf k}+\omega_{\bf p-k})\right)\right]
  \label{1loopint}
\end{equation}
One sees that the one--loop correction is rather complicated and in
general is not equal to $B\lambda n^2\cdot A^{\text{tree}}$, unlike
the cases considered in the body of this paper. The only exception is
the theory in (2+1) dimensions, $d=2$. In that case the integral in
Eq.\ (\ref{1loopint}) is infrared divergent at ${\bf k}=0$ and ${\bf
k}={\bf p}$, the same is true for $B$. If one introduces an infrared
cutoff $p_0$ (say, by considering final particles not exactly at
threshold but with finite momenta of order $p_0$), Eq.\
(\ref{1loopint}) reduces to $A^{\text{1-loop}}=B\lambda n^2\cdot
A^{\text{tree}}$. This result can be easily understood since in (2+1)
dimensions, the loop corrections near threshold are dominated by the
rescattering of final particles. The analysis then is completely
similar to the case of $1\to n$ processes, and further details can be
found in Ref.\ \cite{LRST}.

\newpage

\vspace{3cm}

\newpage

\setlength{\unitlength}{0.5pt}
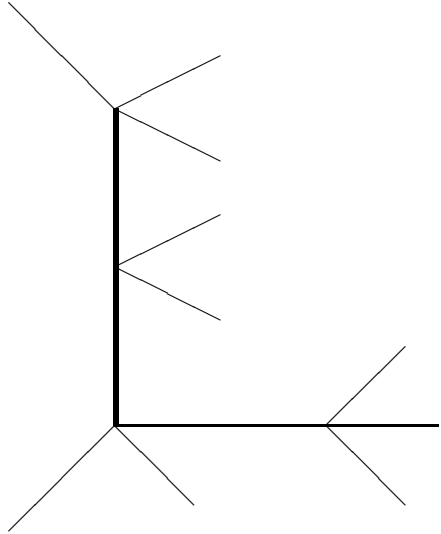
\begin{figure}
\begin{picture}(325,400)(-200,360)
\thicklines
\put(160,680){\line( 0,-1){240}}
\put(162,680){\line( 0,-1){240}}
\thinlines
\put( 80,760){\line( 1,-1){ 80}}
\put( 80,360){\line( 1, 1){ 80}}
\put(160,680){\line( 2, 1){ 80}}
\put(160,680){\line( 2,-1){ 80}}
%\put(160,560){\line( 1, 0){ 80}}
\put(160,560){\line( 2, 1){ 80}}
\put(160,560){\line( 2,-1){ 80}}
\put(160,440){\line( 1, 0){160}}
\put(320,440){\line( 1, 1){ 60}}
\put(320,440){\line( 1, 0){ 85}}
\put(320,440){\line( 1,-1){ 60}}
\put(160,440){\line( 1,-1){ 60}}
\end{picture}
\vspace{1cm}
\caption{A typical tree graph for the process $2\to n$ (ordinary Feynmann
rules are assumed).  When the momentum running along one of the thick lines
is on--shell, the tree amplitude is infinite.} \label{fig:tree2n}
\end{figure}

\vspace{1.25cm}

\setlength{\unitlength}{0.5pt}

\begin{figure}

\begin{picture}(320,105)(-100,700)
\thicklines
\put(100,740){\circle*{10}}
%\put(140,740){\circle*{10}}
\put(180,740){\circle*{10}}
\put(220,800){\circle*{10}}
\put(300,800){\circle*{10}}
\put(340,760){\circle*{10}}
\put(410,760){\circle*{10}}
\put( 80,700){\circle*{ 5}}
\put(480,700){\circle*{ 5}}
\put( 50,695){$\displaystyle{\tau}$}
\put(500,695){$\displaystyle{\tau'}$}
\put( 80,700){\line( 1, 0){400}}
\put(100,740){\line( 1,-1){ 40}}
%\put(140,700){\line( 0, 1){ 40}}
\put(180,740){\line(-1,-1){ 40}}
\put(260,700){\line( 0, 1){ 60}}
\put(260,760){\line(-1, 1){ 40}}
\put(300,800){\line(-1,-1){ 40}}
\put(340,760){\line( 0,-1){ 60}}
\put(410,760){\line( 0,-1){ 60}}

\end{picture}
\vspace{1cm}
\caption{A typical graph in the perturbative expansion of
${\cal D}(\tau,\tau')$}
\label{fig:smallE}
\end{figure}
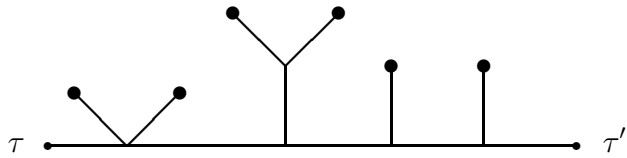
%\end{center}

\vspace{1.25cm}
\setlength{\unitlength}{1pt}
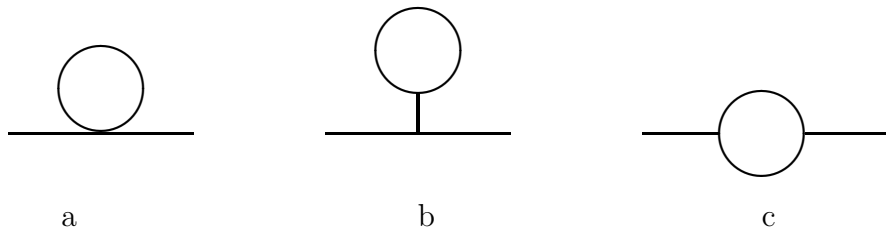
\begin{figure}
\begin{picture}(440,100)(-35,300)
\thicklines
\put(25,350){\line(1,0){70.0}}
\put(60,367){\circle{30}}
\put(45,315){a}

\put(145,350){\line(1,0){70}}
\put(180,365){\line(0,-1){15}}
\put(180,381.5){\circle{30}}
\put(180,315){b}

\put(265,350){\line(1,0){28.5}}
\put(326.5,350){\line(1,0){30}}
\put(310,350){\circle{30}}
\put(310,315){c}

\end{picture}

\vspace{0.5cm}
\caption{The one--loop graphs contributing to ${\cal{D}}(x,y)$}
\label{fig:3diagrams}
\end{figure}

\end{document}